\documentclass[preprint,pre,floats,showkeys,aps,amsmath,amssymb,12pt]{revtex4}
\usepackage{graphicx}
\usepackage{natbib}
\usepackage{amsthm}
\usepackage[left = 1in, top=1.2in,right=1in,bottom=1.2in,a4paper]{geometry}
\usepackage{amsmath}
\usepackage{amssymb}
\usepackage{hyperref}
\newcommand{\ket}[1]{|#1\rangle}
\newcommand{\bra}[1]{\langle #1|}

\usepackage[english]{babel}
\usepackage[utf8]{inputenc}
\usepackage[colorinlistoftodos]{todonotes}

\begin{document}
\title{Tunneling across the Quantum Horizon Does not Resolve the Information Paradox }

\author{Avik Roy\footnote{Email: avikroy@eee.buet.ac.bd}, Moinul Hossain Rahat\footnote{Email: rahat.moin90@gmail.com}}
\affiliation{Department of Electrical \& Electronic Engineering, Bangladesh University of Engineering \& Technology, Dhaka 1000, Bangladesh \vspace{1in}}


\vspace{1in}

\begin{abstract}
Parikh and Wilczek formulated Hawking radiation as quantum tunneling across the event horizon proving the spectrum to be nonthermal. These nonthermality factors emerging due to back reaction effects have been claimed to be responsible for correlations among the emitted quanta. It has been proposed by several authors in literature that these correlations actually carry out information locked in a black hole and hence provide a resolution to the long debated black hole information paradox. This paper demonstrates that this is a fallacious proposition. Finally, it formulates the implications of the no-hair theorem in the context of Parikh-Wilczek spectrum.
\end{abstract}

\maketitle

\section{Introduction}

With Bekenstein's famous thought experiment in \cite{bek-1}, characterizing a black hole as a thermodynamic object has been accelerated in a series of works \cite{penrose},\cite{D.C.},\cite{hawk-0} leading to a complete formulation of black hole thermodynamics \cite{bh-therm}. In his celebrated paper \cite{hawk-1}, Stephen Hawking showed that a black hole of mass $M$ radiates like an ordinary thermodynamic object of temperature $\frac{1}{8\pi M}$. In his subsequnt paper \cite{hawk-2}, Hawking explicitly showed that  black hole radiation is exactly thermal in nature, leading to complete oblivion of the initial configuration that formed the black hole, contrary to what is expected in unitary time evolution of quantum systems. This phenomenon, typically dubbed as the \emph{black hole information paradox}, still remains to be one of the most open problems in research of quantum gravity.

One of the major objections to Hawking's calculations is, despite being a dynamic process, black hole radiation has been analyzed in a fixed space-time background. In this leading order picture, the monotonically increasing entanglement entropy between the black hole interior and exterior gives rise to information loss, reflected in the nonzero value of entanglement entropy (i.e. von Neumann entropy) of the final radiation. It has been speculated that incorporating small quantum gravity effects, back reaction or small perturbations to Schwarzschild geometry might be sufficient to restore unitarity. This issue of \emph{small corrections} was analytically addressed in \cite{peda} and subsequently followed up in \cite{avery},\cite{gidd-shi},\cite{icece},\cite{correlation} to prove that small corrections are insufficient to resolve the information paradox.

However, Parikh and Wilczek successfully incorporated the small back reaction effects during Hawking radiation to correctly estimate the radiation spectrum, proving it to be nonthermal \cite{P-W}. Zhang et al.\! claimed in a series of papers \cite{Zhang-1},\cite{Zhang-2},\cite{Zhang-3},\cite{Zhang-5},\cite{Zhang-4} that small corrections emerging from the nonthermality factors in the transmission amplitude of radiation quanta are sufficient to encode the black hole information as correlations among the radiated quanta. This intrigued a school of thought \cite{Chen},\cite{Singleton},\cite{Israel},\cite{Sakalli} that promoted the results by Zhang et al.\! as a possible resolution of the information loss paradox.

In this paper, we address some major problems regarding the results obtained by Zhang et al. We identify that the analytic expressions used in their work to define and calculate some quantum information theoretic terms are either inconsistent or inadequate. Furthermore, we rigorously formulate that the implications of the no-hair theorem persist even with the nonthermal spectrum of Parikh and Wilczek. This implies that the information paradox cannot be bypassed or resolved by using mere nonthermality of black hole radiation.

This paper is organized as follows. Section \ref{s2} summarizes the leading order calculation of Hawking radiation as presented in \cite{exactly},\cite{peda},\cite{isnot}. Section \ref{s3} provides a brief review of the results by Parikh and Wilzeck in \cite{P-W} and how these results have been interpreted as a resolution of the information paradox in \cite{Zhang-1}. Section \ref{s4} presents the main results of this paper and contains the detailed arguments identifying the problems of the results outlined by Zhang et al.\! and others. Section \ref{s5} concludes by summarizing the key results of this paper and their implications.

\section{Information Loss in Hawking Radiation: Leading Order Results}\label{s2}

Hawking radiation can be visualized as pair production near the black hole horizon. The joint system of the particle pair near the horizon can be given as \cite{hawk-2},\cite{gidd} --
\begin{equation}\label{pair}
\ket{\Psi}_{pair} = Ce^{\beta c^{\dagger} b^{\dagger}}\ket{0}_c \ket{0}_b
\end{equation}
where $\beta$ is a number of order unity, $c^\dagger$ and $b^\dagger$ are creation operators, $\ket{0}$ represents the vacuum state and $c,b$ represent the ingoing and outgoing quanta respectively. It is trivial to note that $b$ and $c$ states are highly entangled. This nature of entanglement is crucial to ensure that the horizon remains an innocuous place for an infalling observer, as exacted by the equivalence principle. In fact, the time reversed infinite blueshift factor due to the outgoing particles is perfectly canceled by the entangled ingoing partner. Being crucial for the black hole geometry, it is this entanglement that gives rise to the information loss paradox. We shall pursue the leading order analysis by making the following assumptions--

(a) As described in \cite{exactly}, \cite{peda}, the geometry of spacetime is foliated by a set of low curvature spacelike `nice' slices that are regular at the horizon. Local quantum field theory describes the essential physics of quantum evolution over the slices.

(b) $\ket{\Psi}_M$ represents the state vector of the initial matter configuration composing the black hole by gravitational collapse.

(c) Instead of the infinite dimensional state vector (\ref{pair}), the newly evolved pair is approximated as
\begin{equation}\label{LO}
  \ket{\Psi}_{pair} = \frac{1}{\sqrt{2}}\left(\ket{0}_c\ket{0}_b + \ket{1}_c\ket{1}_b\right)
\end{equation}
 on a much simpler qubit space. The entropy of entanglement associated to the outgoing quantum is given by
\begin{equation}
  S_{ent} = \ln 2
\end{equation}

(d) At each successive step, stretched spacelike slice causes a new pair to evolve according to (\ref{LO}) while the earlier qubits and the matter state moves farther along the spacelike slice. The geometry of the nice slices near the horizon remains the same as before. By the \emph{no-hair} postulate, the black hole solution bears no imprint of earlier radiation.  Hence, the effects of earlier quanta can be ignored to give the joint state after $N$ pairs have evolved. This gives the joint state
\begin{equation}\
  \ket{\Psi} = \ket{\Psi}_M \otimes\ket{\Psi}_1 \otimes \ket{\Psi}_2 \otimes \ldots \otimes \ket{\Psi}_N
\end{equation}
where
\begin{equation}
\ket{\Psi}_i = \frac{1}{\sqrt{2}}\left(\ket{0}_{c_i}\ket{0}_{b_i} + \ket{1}_{c_i}\ket{1}_{b_i}\right)
\end{equation}
Hence, the entanglement entropy after $N$ emissions is given by
 \begin{equation}
  S_{ent} = N\ln 2
\end{equation}
Unless we are left with a remnant of Planck scale, this monotonically rising entanglement entropy implies a mixed state description for the radiation in spite of the black hole being formed from the gravitational collapse of a matter configuration in pure state $\ket{\Psi}_M$. Therefore, any resolution of the information paradox will have to advocate a way of bypassing this conundrum of monotonically increasing entanglement entropy.

A parametric method of incorporating and quantifying deviations from the leading order results has been championed by Mathur in \cite{peda}. With an abstract mathematical formalism, he proved that small corrections do not suffice to resist the monotonic rise of entanglement entropy. It should be noted that neither \cite{peda} nor any of its follow ups in \cite{avery},\cite{correlation} sticks to some particular physics of small correction. Results in these papers imply that leading order formulation needs to be modified by order unity to restore unitarity in black hole radiation. As pointed out in \cite{subadd}, unitarizing Hawking radiation will call for some novel physics if the implications of black hole geometry are accepted. Otherwise, a way must be identified to bypass the no-hair theorem.

\section{Resolution of the Information Paradox as interpreted by Zhang et al.}\label{s3}

The continuously rising entanglement entropy as discussed in the earlier section has often been attributed to the thermal nature of the black hole radiation. With a fixed black hole geometry, the created pairs are always uncorrelated and hence state of the evolved pair can be approximated as (\ref{LO}). Parikh and Wilczek derived the spectrum of black hole radiation considering the effect of back reaction on the background geometry and strictly imposing the law of conservation of energy \cite{P-W}. In the picture they conceived, particle pairs are created just behind the horizon and one partner can tunnel across the quantum horizon to materialize as a real particle. The probability of tunneling for a particle with energy $E$ from a black hole of mass $M$ is given by
\begin{equation}\label{tunnel}
\Gamma (E) \sim \exp\left[ -8\pi E \left( M - \frac{E}{2} \right)  \right] = e^{\Delta S}
\end{equation}
The last part of the equation comes from recognizing the change in Bekenstein-Hawking entropy $\Delta S = 4\pi (M-E)^2 - 4\pi M^2   = -8 \pi E \left( M - \frac{E}{2}\right)$. The tunneling probability (\ref{tunnel}) calculated by Parikh and Wilczek is clearly distinct from the thermal distribution calculated by Hawking as $\Gamma (E) = \exp \left( -8\pi ME \right)$.

Implications of this nonthermal distribution in terms of correlations among the radiated particles have been investigated in \cite{Parikh}, \cite{Arzano}; but it was Zhang et al.\! who cleverly recognized the nontrivial correlations among the radiated quanta \cite{Zhang-1},\cite{Zhang-2}. Their findings can be summarized in the following two points. Firstly, the event of emission of the $i$-th particle with energy $E_i$ is statistically dependent on the events of earlier emissions, i.e.
\begin{equation}\label{Z1}
\Gamma (E_i) \ne \Gamma (E_i| E_1, E_2, \ldots E_{i-1})
\end{equation}

Secondly, the total \emph{entropy} of the radiation subsystem at any stage of evaporation is equal to the change in the Bekenstein-Hawking entropy of the black hole.
\begin{equation}\label{Z2}
S(E_1, E_2, \ldots E_k) = \sum_{i=1}^{k} S (E_i | E_1, E_2, \ldots E_{i-1}) = 4 \pi \left[  M^2 - \left( M -  \sum_{i=1}^{k} E_i\right)^2 \right]
\end{equation}
Hence, after the entire black hole has been evaporated, the entropy of the radiation is equal to the original Bekenstein-Hawking entropy $S_{BH}$ of the black hole. After $n$ particles with energies $(E_1, E_2, \ldots E_n)$, subject to the constraint of energy conservation $\sum_{i=1}^{n} {E_i} = M$, have been emitted, (\ref{Z2}) reduces to
\begin{equation}\label{Z3}
S(E_1, E_2, \ldots E_n) = 4 \pi M^2 = S_{BH}
\end{equation}
which has been interpreted as \emph{conservation of entropy}. Zhang et al.\! identify (\ref{Z3}) as the analytic expression for the resolution of the black hole information paradox. The same line of thought has been pursued in \cite{Chen},\cite{Singleton},\cite{Israel},\cite{Sakalli}. Their arguments can be summarized as follows.

Firstly, the equality of Bekenstein-Hawking entropy and radiation entropy implies that radiation carries away all information that is locked inside the black hole. As argued in \cite{Zhang-1}, emission of the $i$-th particle with an energy $E_i$ reveals $S(E_i|E_1 , E_2, \ldots E_{i-1}) = -\ln \Gamma(E_i|E_1 , E_2, \ldots E_{i-1})$  amount of information. Total information carried out when the entire black hole has been evaporated is given by (\ref{Z3}). This information is carried away by the nontrivial  correlations among the radiated quanta.

Secondly, by virtue of the entropic equality, both black hole and the radiation can be identified with $e^{S_{BH}}$ microstates.  Hence it implies a one to one correspondence among the black hole microstates and the different radiation configurations which inherently can be interpreted as a unitary matrix \cite{Singleton}. Thirdly, black hole entropy accounts for the different radiation configurations, i.e., the distribution of energy among the radiated quanta\cite{Israel}.

However, as it will be showed in the following section, these ideas are at odds with the idea of conservation of quantum information in a unitary evolution.

\section{Difficulties with Interpretation by Zhang et al.}\label{s4}

Premier objections against the propositions of \cite{Zhang-1} were raised by Mathur in \cite{isnot} and Medved et al. in \cite{Medved}. It was correctly addressed by Mathur that their analysis does not recognize the issue of continuously rising entanglement entropy. Perhaps, the missing prefactor in (\ref{tunnel}) contains the essential physics of entanglement across the horizon. According to Mathur, the missing prefactor in (\ref{tunnel}) is of significant physical importance because the quantities on both sides of this relation have different units. Hence, it is the missing prefactor that corresponds to the underlying physics of tunneling across the horizon.

However, even with the missing prefactor, resolution of the information paradox in the tunneling picture is a misleading paradigm. The following subsections identify the major problems with such ideas.

\subsection{A misleading definition of \emph{entropy}}
Let us first note that the definition of entropy repeatedly used by Zhang et al.\! is actually a misleading measure of quantum information. They address that \emph{entropy} associated with the emission of a particle of energy $E$ is given by
\begin{equation}\label{zh-ent}
S (E) = -\ln \Gamma (E)
\end{equation}
This quantity, which reduces to the value of $8\pi E \left( M - \frac{E}{2} \right)$ under the approximation of equality in (\ref{tunnel}), has been attributed as the measure of \emph{quantum information} by Zhang et al.\! and others who followed them.

Such an attribution is not  technically correct. The information paradox is an explicit violation of unitarity which is analytically expressed as the conservation of von Neumann entropy. This quantity for a quantum system described by the density operator $\hat{\rho}$ is given by
\begin{equation}\label{VN}
S_{vN}(\hat{\rho}) = - \mathrm{Tr}\left(\hat{\rho} \ln \hat{\rho}\right)
\end{equation}

There is no straightforward relation between the quantities in the last two equations. Von Neumann entropy is a basis independent measure whereas the other expression in context considers probabilities associated with the measurement of energy of the emitted particle. The pure or mixed nature of a quantum state could be irrelevant in quantifying (\ref{zh-ent}). On the other hand, von Neumann entropy of a radiated quantum completely depends on the nature of entanglement between the quantum and the remaining black hole. Parikh-Wilczek derivation of nonthermal black hole spectrum alone does not provide any Hilbert space description of the radiation quanta. Hence, it would be problematic to deduce that conservation of the quantity in (\ref{zh-ent}) or any of its derivatives like in (\ref{Z3}) is equivalent to conservation of unitarity.

\subsection{The erroneous mutual information}

In this subsection, we shall prove that Zhang et al.\!'s calculation of mutual information in \cite{Zhang-1} is erroneous. Mutual information between two successive emissions of energies $E_1$ and $E_2$ is given as
\begin{equation}\label{zh-mu}
S(E_2 : E_1) = -\ln\Gamma(E_2) + \ln\Gamma(E_2|E_1)
\end{equation}
This quantity does not measure quantum correlations between the first two quanta. It quantifies the amount of correlation between the outcomes of two successive measurements and depends on the choice of measurement basis. It is trivial to note that a nonzero value of this quantity indeed implies some sort of quantum correlation between the particles identified with energies $E_1$ and $E_2$. This is simply because   outcome of a measurement on a subspace is trivially independent of the measurement results for some other subspace if these subspaces are uncorrelated i.e. unentangled. But there is no definite way to determine how to identify a nonzero value of (\ref{zh-mu}) as a quantitative measure of quantum mutual information.

In fact (\ref{zh-mu}) is not a reliable measure of quantum mutual information and depends on the choice of the basis of measurement. A simple example can demonstrate this. Suppose the joint state of two particles is given by the qubit state
\begin{equation}\label{st-1}
\ket{\Psi}_{AB} = \frac{1}{\sqrt{2}}\ket{0}_A \otimes\frac{1}{\sqrt{2}} \left(\ket{0}_B + \ket{1}_B\right) + \frac{1}{\sqrt{2}}\ket{1}_A \otimes\frac{1}{\sqrt{2}} \left(\ket{0}_B - \ket{1}_B\right)
\end{equation}
Clearly, the particles denoted by $A$ and $B$ are maximally entangled and this fact should be reflected in any appropriate measure of quantum correlations between them. However, if the quantity in (\ref{zh-mu}) is calculated for the aforementioned system, it can be easily deduced that
\begin{equation}
S(a : b) = 0 \qquad\qquad \forall a,b \in \{0, 1\}
\end{equation}
This could be interpreted as statistical independence of the subsystems $A$ and $B$, which would be an erroneous deduction. In fact, this result actually identifies that measurement outcomes for these two subsystems in a certain basis are statistically uncorrelated. Hence, it would be incorrect to make any conclusion about the nature of Hawking radiation depending on ambiguous and basis dependent definitions like (\ref{zh-ent}) or (\ref{zh-mu}).

Thus far we have identified the problems with the measure of correlations used by Zhang et al. Let us now prove that the so called conservation of entropy in (\ref{Z3}) does not imply conservation of quantum information. Assume a nonunitary evolution of the pure state in (\ref{st-1}) given by
\begin{equation}
\ket{\Psi}_{AB} \to \hat{\rho}_{AB}
\end{equation}
where $\hat{\rho}_{AB}$ is a density matrix denoting the mixed state
\begin{equation}
\hat{\rho}_{AB} = \frac{1}{2} \left(\ket{0}\bra{0} + \ket{1}\bra{1}\right)_A \otimes \frac{1}{2} \left(\ket{0}\bra{0} + \ket{1}\bra{1}\right)_B
\end{equation}
A quantity equivalent to the Bekenstein-Hawking entropy of the black hole associated with $\ket{\Psi}_{AB}$ should be the logarithm of the dimension of the Hilbert space describing the joint system $AB$. In fact, logarithm of the dimension of the Hilbert space associated with a given quantum system has been identified as the thermodynamic entropy of that system by Page in \cite{Page}. Bekenstein-Hawking entropy is indeed the canonical measure of thermodynamic entropy of a black hole. Stretching this resemblance to the toy model of a 2 qubit black hole in (\ref{st-1}), the value of Bekenstein-like entropy for this system is given by
\begin{equation}\label{false1}
S_{BH}^{\Psi} = 2 \ln 2
\end{equation}

Let us now calculate the amount of information conveyed by a certain measurement outcome of the radiation state $\hat{\rho}_{AB}$. It is easy to show that
\begin{align}\label{false2}
S(a, b) &= S(a) + S(b|a) \notag \\
& = 2 \ln 2 \qquad\qquad\qquad \forall a,b \in \{0, 1\}
\end{align}
The equality of (\ref{false1}) and (\ref{false2}) is an essential reproduction of the same result as (\ref{Z3}), but as it is evident this is inadequate to claim unitarity in black hole evaporation.

It could be now safely concluded that the quantitative measures explored in \cite{Zhang-1} are inconsistent with the traditional understanding of quantum information and at least inadequate to address the core difficulty about the black hole information paradox. Instead of using simple toy models, the following subsection explicitly engages with the calculations based on the Parikh-Wilczek spectrum of black hole radiation and shows why and how it is subject to the same problem as the Hawking spectrum.

\subsection{Implications of the no-hair theorem}

Let us now turn to the interpretation of black hole entropy. According to Zhang et al.\! black hole entropy accounts for the different possible radiation configurations, i.e., the distribution of black hole energy among the radiated particles $(E_1, E_2 \ldots E_n)$. By virtue of  (\ref{tunnel}) and (\ref{Z3}),
\begin{equation}\label{Z4}
\Gamma(E_1, E_2 \ldots E_n) = \exp[-S_{BH}]
\end{equation}
From the fundamental assumption of statistical physics that all possible microstates are equally likely, the number of microstates associated with the radiation is given by $\Omega = e^{S_{BH}}$, the same as the number of microstates of the initial black hole.

This interpretation is fallacious. First, let us note the Parikh-Wilczek formula for tunneling probabilities. Despite being nonthermal, it is subject to the same intrinsic feature of Hawking spectrum that eventually leads to the problem of information loss -- the no-hair theorem. For a black hole of mass $M$, irrespective of the (quantum) state of the collapsed matter, the tunneling probabilities in (\ref{tunnel}) are solely expressed in terms of $M$. This implies that the no-hair theorem persists even with a nonthermal spectrum of black hole radiation. Not only the space-time geometry is independent of the initial matter state collapsing to form the black hole, but also the back reaction effects are identical. As a result, though incorporating such effects results in deviation from strict thermality of Hawking radiation, such effects actually cannot relay any quantum information. We shall now prove this statement rigorously.

Let us identify a Hilbert space of dimension $e^{S_{BH}}$ associated with a black hole of mass $M$ (and hence Bekenstein-Hawking entropy $S_{BH} \sim M^2$) spanned by a finite set of basis vectors $\{\ket{\psi_i}\}$. After the black hole has completely evaporated, the quantum state of the radiation also has to be described by a Hilbert space of the same dimension and hence must be spanned by another set of vectors $\{\ket{\lambda_i}\}$. If the distinct $e^{S_{BH}}$ modes of evaporation truly convey the quantum information of the black hole state, then these modes should be identified as a complete set of basis vectors spanning the radiation space. Let us identify these modes as the $\ket{\lambda_i}$ states, i.e.
\begin{equation}
\ket{\lambda_i} = \ket{E_{i_1}, E_{i_2}, \ldots E_{i_{n_i}}}
\end{equation}

Let black hole evaporation be described by a unitary matrix $\hat{U}$. It is then always possible to make a choice of the basis vectors $\ket{\psi_i}$ so that
\begin{equation}\label{U}
\hat{U}\ket{\psi_i}=\ket{\lambda_i}
\end{equation}
Now, a generic black hole state is described by the density matrix $\hat{\rho}_{BH}$ as
\begin{equation}
\hat{\rho}_{BH}=\sum_{i,j}{C_{ij}\ket{\psi_i}\bra{\psi_j}}
\end{equation}
where $\sum_i C_{ii} = 1$. After the black hole has been evaporated entirely, the density matrix representation of the radiation system is given by
\begin{align}\label{rho}
\hat{\rho}_{rad} = \hat{U}\hat{\rho}_{BH}\hat{U}^\dagger = \sum_{i,j}{C_{ij}\ket{\lambda_i}\bra{\lambda_j}}
\end{align}
It should be noted that this form of evolution implies that $\hat{U}$ acts on the black hole interior alone. However, it has been argued by some authors \cite{verlinde} that a black hole, soon after its formation, should be entangled with its immediate environment. When such a black hole is completely evaporated, its surrounding environment should be left in a unique pure state (the vacuum state). This suggests that (i) $\hat{U}$ should act on both the interior and the environment and (ii) the radiation should contain extra excitations. These extra excitations correspond to the degrees of freedom of the evironment. Hence, the radiation space should be described by a Hilbert space of dimension larger than $e^{S_{BH}}$. If (ii) is true, the Zhang et al. \! formalism then breaks down automatically. Therefore, for the purpose of the present work, \ref{rho} is sufficient.

Let us now assume that the radiation is measured in the $\ket{\lambda_i}$ basis. The probability that the mode of evaporation is found in the $\ket{\lambda_i}$ state is given by
\begin{align}\label{calc}
\mathrm{Pr}(E_{i_1}, E_{i_2}, \ldots E_{i_{n_i}}) & =\sum_j \mathrm{Pr}(\lambda_i|\psi_j)\mathrm{Pr}(\psi_j) \notag \\
& = \sum_j \delta_{ij}\bra{\psi_j}\hat{\rho}_{BH}\ket{\psi_j} \notag \\
& = \sum_j \delta_{ij}C_{jj} \notag \\
& = C_{ii}
\end{align}
In the second line, $\mathrm{Pr}(\lambda_i|\psi_j) = \delta_{ij}$ by virtue of equation (\ref{U}). From \ref{Z4} and (\ref{calc}),
\begin{equation}\label{calc2}
C_{ii} = \exp(-S_{BH})
\end{equation}
This relation is subject to the same difficulty as the information paradox. Where the quantity in the left is completely state dependent, the one in the right is independent of the initial black hole state. So modes of evaporation cannot actually account for the black hole entropy since the result in (\ref{Z3}), owing to the no-hair theorem, does not make any inference about the state of the matter that formed the black hole. It essentially states that every mode of evaporation is equally likely independent of the black hole state $\ket{\psi_i}$ and hence the conundrum of (\ref{calc2}) arises.

In fact, a true account of black hole information calls for a Hilbert space description of the black hole radiation. As it has been illustrated in this paper, the Parikh-Wilczek spectrum of black hole radiation alone is not sufficient to manifest a unitary description of the evaporation process. However, it has been showed by Braunstein and Patra in \cite{brauns} that a Hilbert space description of black hole evaporation conforms to the Parikh-Wilczek spectrum in (\ref{tunnel}). Their results put forward two significant ideas in the context of our paper.

First, we can attribute a proper interpretation to the quantity in (\ref{zh-ent}) as the thermodynamic entropy of the radiation. Zhang et al.\! actually proved that black hole evaporation preserves thermodynamic entropy.

Second, \cite{brauns} proves that (i) (\ref{tunnel}) holds irrespective of the details of the underlying unitary process and (ii) two black holes with identical mass will have identical spectrum, even for a unitary theory. This suggests that the idea of \emph{no hair} persists with the Parikh-Wilczek results. The nonthermal spectrum alone does not preserve any details of the microscopic structure. It also fails to prescribe any unitary mechanism for black hole evaporation. Therefore, the conclusions by Zhang et al.\! remain problematic.

\section{Conclusion}\label{s5}
Parikh-Wilczek spectrum for black hole radiation correctly incorporates the effect of back reaction and hence establishes its nonthermal nature. This nonthermality was attributed to be the source of the necessary correlations among the radiated quanta required to resolve the information paradox. This paper demonstrates the problems intrinsic to this proposition and why this resolution fails to address the premier issue of unitarity. Identifying that Parikh-Wilczek spectrum is subject to the same implications of the no-hair theorem as Hawking spectrum, we conclude that deviation from thermality alone is hardly adequate to restore unitarity in the process of black hole evaporation. Rather, if one sticks to the physics of local quantum field theory, black hole information paradox can only be avoided by addressing a way of bypassing the no-hair theorem.

\section*{Acknowledgement}
The authors would like to thank Samuel Braunstein and Mahbub Majumdar for useful discussions and helpful comments.
\bibliography{tunnel2}
\bibliographystyle{unsrt}

\end{document}